\documentclass[%
 reprint,
 amsmath,amssymb,
 aps,
]{revtex4-1}

\usepackage{graphicx}% Include figure files
\usepackage{dcolumn}% Align table columns on decimal point
\usepackage{bm}% bold math
\usepackage{placeins}
\usepackage{xcolor}

\begin{document}

\preprint{APS/123-QED}

\title{Effect of Crab cavity HOMs on the coupled-bunch stability of HL-LHC}% Force line breaks with \\
%\thanks{This research is supported by the High-Luminosity LHC project.}%

\author{S. A. Antipov}
 \email{Sergey.Antipov@cern.ch}%
\author{N. Biancacci}%
\author{J. Komppula}%
\author{E. M\'etral}%
\author{B. Salvant}%
 
\affiliation{%
 CERN, Geneva 1211, Switzerland
}%

\author{A. Burov}
 \affiliation{
 Fermilab, Batavia, IL 60510, USA
}%

\date{\today}% It is always \today, today,

\begin{abstract}
High frequency High Order Modes can significantly affect the coupled-bunch stability in a circular accelerator. With a large enough shunt impedance they may drive a coupled-bunch transverse instability. We have developed an analytical model and implemented it in the NHT Vlasov solver. The results show that the instability is characterized by the excitation of many azimuthal intra-bunch modes, which have similar growth rates, that make the traditional remedies such as a flat resistive feedback and chromaticity inefficient in suppressing it. For the High Luminosity Large Hadron Collider this could result in a significant increase of the stabilizing Landau octupole current, up to $\sim 100$~A ($\sim 20\%$ of the maximum available current). In order to limit the increase below 10~A ($\sim 2\%$), the transverse shunt impedance has to be kept below 1 M$\Omega$/m.
\end{abstract}

\pacs{Valid PACS appear here}% PACS, the Physics and Astronomy
\maketitle

%\tableofcontents

\section{\label{sec:Intro} Introduction}

Crab cavities are a key technology in many future colliders. They allow increasing the luminosity at a given beam crossing angle by rotating the bunches in the plane of collision and restoring an effective head-on collision at the interaction point \citep{bib:CC_Palmer, bib:CC_Oide, bib:CC_paper}. The crab crossing was first implemented in the electron-positron circular collider KEK-B, where it allowed increasing the luminosity and beam-beam tune shift \citep{bib:CC_KEK}. Since then crab cavities were proposed for many potential future high energy machines. They could re-establish head-on collisions in linear electron-positron colliders such as Compact LInear Collider (CLIC) and International Linear Collider (ILC) \citep{bib:Verdu-Andres} and reduce beam-beam effects, allowing increasing the luminosity, in circular hardon colliders such as FCC-hh \citep{bib:FCC_CDR} and High Energy Large Collider (HE-LHC) \citep{bib:HE-LHC_CDR} or electron-ion rings such as Jefferson Lab Electron Ion Collider (JLEIC) and electron-Relativistic Ion Collider (eRHIC) \citep{bib:Verdu-Andres,bib:Verdu-Andres_2017}.

Large Hardon Collider (LHC) also plans installing the cavities at two interaction points as a part of its High Luminosity (HL) upgrade \citep{bib:HL-LHC_TDR}. The crab cavities will allow increasing its virtual luminosity by a factor two, making it possible to extend the luminosity levelling at $5.0\times 10^{34}$~cm$^{-2}$s$^{-1}$ in the Nominal operational scenario or at $7.5\times 10^{34}$~cm$^{-2}$s$^{-1}$ in the Ultimate. Overall, considering realistic operational scenarios, usage of crab cavities reduces the effective event pile-up density by around 20\% and increases the integrated luminosity by about 10\% \citep{bib:Rogelio_Madrid}.

\begin{figure}[h]
\centering
\includegraphics[width = 2.3in]{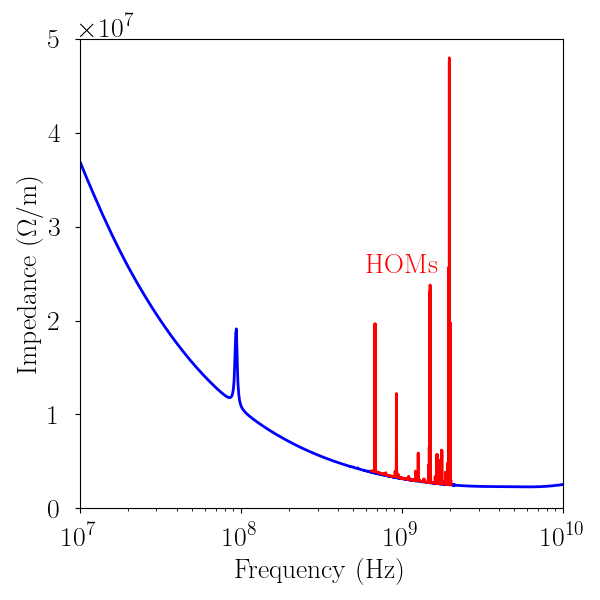}
\caption{
Higher order modes of the crab cavities (shown in red) dominate the HL-LHC transverse impedance (blue line) at the frequencies around 1~GHz. Real part of the horizontal dipolar impedance is plotted vs frequency for the ultimate scenario at the top energy of $E=7$~TeV. $\beta^* = 41$~cm, 4 cavities per IP and beam.
}
\label{fig:problem}
\end{figure}

\begin{figure}
  \centering
  \includegraphics[width = 2.32in]{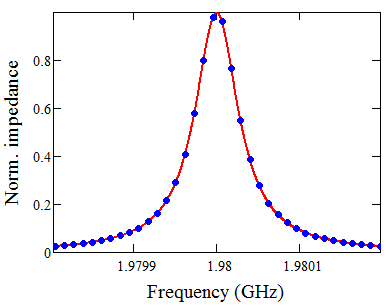}
  \caption{
  A typical crab HOM (red line represents the real part of its normalized impedance) covers multiple coupled-bunch frequency lines (blue dots) and can therefore drive multiple coupled-bunch modes.
	}
  \label{fig:driv_modes}
\end{figure}

Crab cavities usually have multiple transverse higher order modes (HOMs), that start dominating the high-frequency part of impedance as the beta-function at the interaction point $\beta^*$ is squeezed down (Fig.~\ref{fig:problem}). A typical HOM has a width much greater than the revolution frequency (for example, about 1~MHz vs 11~kHz for High-Luminosity LHC) and thus drives several coherent coupled-bunch modes (Fig.~\ref{fig:driv_modes}). Thus, if the mode's shunt impedance is large enough it might have a significant impact on the coupled-bunch beam stability, which is crucially important for circular colliders where the beam typically has to be stored millions of revolutions before collision.

A simulation of the complete coupled-bunch problem is computationally rather demanding. For example, for HL-LHC one would need to track 2760 bunches with between $10^4$ and $10^6$ macroparticles each for $10^5 - 10^6$ turns depending on the study. Therefore a Vlasov solver that reduces the complexity of the task to a standard eigenvalue problem seems a better alternative. The Nested Head-Tail (NHT) Vlasov solver \citep{bib:NHT} offers an efficient way to treat the multi-bunch problem, but the coupled-bunch approximation it has employed so far cannot be applied to the high frequency HOMs. We have extended the NHT code to include an exact solution of the coupled-bunch problem and used it to study the effect of crab cavity HOMs on the transverse beam stability in HL-LHC. Our goal was to, first, quantify the effect of crab cavity HOMs in terms of additional stabilizing octupole current, and second, establish the limits on the acceptable HOM shunt impedance.

This paper is organized in the following way: in Sec.~II we follow the \textit{nested head-tail} approach of Ref.~\citep{bib:NHT} and derive the linear model of coupled-bunch transverse stability from the Vlasov equation. We skip some details of the derivation, which can be found in \citep{bib:NHT,bib:Chao}.

In Sec.~III we extend the formalism to the case of long-range high-frequency wakes. The new parts (to our knowledge) of the analysis include the explicit treatment of the resulting coupled-bunch and intra-bunch modes for an arbitrary intensity and longitudinal beam distribution. Using this approach we show that due to the excitation of a large number of azimuthal intra-bunch modes neither damper, nor chromaticity play a significant role in the dynamics for sufficiently high HOM frequencies. Then we compare the results with the known formulas for the flat wake case and the weak head-tail limit and with the results of multi-bunch tracking simulations.

In Sec.~IV we apply our model to compute the stability estimates for HL-LHC crab cavity HOMs. Following the approach of previous studies \citep{bib:Nicolo_IPAC,bib:Sergey_Madrid}, we quantify the effect of crab HOM in terms of extra stabilizing octupole current for different operational scenarios.

\section{\label{sec:NHT}Nested Head-Tail Formalism}

We consider the transverse modes of a coherent motion of a proton beam, following the approach of \citep{bib:Chao}. Assuming no significant diffusion or external damping effects, such as synchrotron radiation, and no coupling between the two transverse planes, the evolution of the beam in the phase space can be described by a Vlasov equation:
\begin{equation} \label{eq:1}
	\frac{d\psi}{ds} = 0,
\end{equation}
where $\psi = \psi(y, p_y, z, \delta; s)$ is the distribution function, depending on the transverse $(y, p_y)$ and longitudinal $(z, \delta)$ positions and momenta and the azimuthal position around the ring $s$. It is helpful to express Eq.~(1) in the polar coordinates $(y, p_y)\rightarrow(q, \theta)$, $(z, \delta)\rightarrow(r, \phi)$:
\begin{equation} \label{eq:2}
	\frac{\partial\psi}{\partial s} + \frac{\omega_\beta}{c}\frac{\partial\psi}{\partial\theta} + \frac{F_y(z,s)}{E}\frac{\partial\psi}{\partial p_y} + \frac{\omega_s}{c}\frac{\partial\psi}{\partial\phi} = 0,
\end{equation}
where $\omega_{\beta,s}$ are the betatron and synchrotron frequencies, $E$ is the beam energy, and $F_y$ is the transverse wake force generated by the dipole moment of the beam, and $c$ is the speed of light. Here and further we consider a relativistic beam, i.e. its velocity $v \approx c$. Further assuming a small perturbation that describes a dipole oscillation mode on top of the unperturbed distribution, which is independent of the angular variables, $\psi_0(q,r) = f_0(q)g_0(r)$ we can write
\begin{equation} \label{eq:3}
	\psi = \psi_0(q,r) + f_1(q,\theta)g_1(r,\phi)e^{-i\Omega s/c},
\end{equation}
where $\Omega$ is the frequency of the mode and $f_1$ and $g_1$ describe its the transverse and longitudinal structure. Accounting for the chromatic dependence of the betatron frequency, Eq.~(\ref{eq:3}) can then be linearized with respect to the perturbation

\begin{equation} \label{eq:Linearized_Vlasov}
	\begin{split}
		& (-i\frac{\Omega}{c}f_1g_1 + \frac{\omega_\beta}{c}(1+\xi\delta)\frac{\partial f_1}{\partial \theta}g_1 + \frac{\omega_s}{c}f_1\frac{\partial g_1}{\partial \phi})\times\\
        & \times e^{-i\Omega s/c} - \frac{F_y c}{E \omega_\beta}\sin(\theta)f_0'g_0 = 0,
	\end{split}
\end{equation}
where $\xi$ stands for chromaticity. Searching for the self-consistent solutions of (\ref{eq:Linearized_Vlasov}) for $\Omega$, $f_1$, and $g_1$ one eventually (after the separation of the transverse and the longitudinal variables) arrives at a system of integral equations on the longitudinal degree of freedom
\begin{widetext}
\begin{equation} \label{eq:BIG_AND_SCARY}
	(\Omega - \omega_\beta -l\omega_s)x_l R_l(r) = -i\frac{\pi r_0\omega_s}{\gamma\omega_\beta T_0^2\eta}g_0(r)\sum_{l'=-	\infty}^{+\infty}\int_0^\infty r'dr'x_{l'}R_{l'}(r')i^{l-l'} \sum_{p=-\infty}^{+\infty}Z_1^\perp(\omega') J_l((\omega'-\omega_	\xi)\frac{r}{c}) J_{l'}((\omega'-\omega_\xi)\frac{r'}{c}),
\end{equation}
\end{widetext}
where $Z_1^\perp$ is the transverse dipole impedance of the machine sampled at the frequencies $\omega' = p\omega_0 + \omega_\beta + l\omega_s$, $J_l$ is the $l$-th order Bessel function of the first kind, $r_0$ - classical proton radius, $T_0$ - revolution period, $\eta$ is the slippage factor, and $\omega_\xi = \xi\omega_0/\eta$ is the chromatic angular frequency shift. The longitudinal structure of the mode is represented as a Fourier series:
\begin{equation} \label{eq:Fourier_decomposition}
g_1(r,\phi) = \sum_{l = -\infty}^{+\infty}x_l R_l(r)e^{il\phi}e^{i\omega_\xi z/c},
\end{equation}
where $l$ is the azimuthal mode number and $R_l(r)$ - the corresponding radial function.

\begin{figure}
  \centering
  \includegraphics[width = 3.5in]{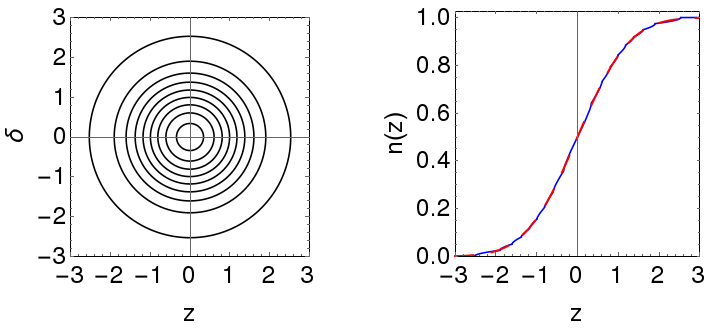}
  \caption{
  Longitudinal phase space distribution of the nested airbag beam (left) and the corresponding normalized integrated charge density (right): the blue line depicts the integrated density for $n_r = 9$ air-bags, the dashed red - for a Gaussian beam.
	}
  \label{fig:Rings}
\end{figure}

The system (\ref{eq:BIG_AND_SCARY}) can be elegantly solved if one models the longitudinal distribution by a set of equally populated nested air-bag rings, as proposed in Ref.~\citep{bib:NHT} (Fig.~\ref{fig:Rings}). The action space is divided in $n_r$, such that each slice $I_n$ contains the same number of particles. A mean action on each slice defines the radius of the corresponding ring $r_n = c \tau_n$. Then the radial distribution function can be approximated as:
\begin{equation} \label{eq:Basis_functions}
g_0(r) = \sum_{n = 1}^{n_r}\frac{N\eta}{2\pi\omega_s n_r\tau_n}\delta(r-c\tau_n),
\end{equation}
where $N$ is the number of particles in a bunch. Then for each ring all radial functions $R_l(r)$ degenerate into delta-functions $\delta(r-c\tau_n)$ and the system of integral equations reduces to a system of linear equations on the coefficients $x_l$. The Bessel functions in (\ref{eq:BIG_AND_SCARY}) need to be computed only at a discreet set of values $\chi_n^{\prime} = (\omega^{\prime} - \omega_\xi)\tau_n = \omega^{\prime}\tau_n - \chi_n$, where $\chi_n$ stands for the conventional head-tail phase advance of the ring $n$.

In order to correctly treat the initial eigenvalue problem the number of rings has to reflect the wake properties. For the LHC wake, dominated by the resistive wall of its collimation system, a choice of $n_r = 5$ 
works reasonably well \citep{bib:NHT}. If the wake includes prominent high-frequency components the number of rings has to be increased such that the characteristic phase advance between two neighbors is small:
\begin{equation} \label{eq:crit}
	2\pi f\Delta \tau \ll 1,
\end{equation}
where $f$ is the characteristic frequency of the wake, and $\Delta\tau$ is the distance between the rings.

For the case of a single air-bag of a radius $\tau$ we obtain a simple eigenvalue problem \citep{bib:Chao}:
\begin{equation} \label{eq:NOT_SO_BIG_AND_SCARY}
  \frac{\Delta\omega}{\omega_s}x_l = l x_l -i\kappa\sum_{l',p = -\infty}^{+\infty} x_{l'}i^{l-l'} Z_1^\perp(\omega') J_l(\chi') J_{l'}(\chi'),
\end{equation}
where $\Delta\omega = \Omega - \omega_\beta$ is the complex frequency shift, $\kappa = \frac{N r_0 c}{2\gamma\omega_\beta\omega_s T_0^2}$ is the normalized intensity parameter, and $\chi = \omega_\xi \tau$ is the head-tail phase advance. 
It is worth mentioning that already this simple and apparently not realistic approximation already incorporates all the Physics of the problem and is a powerful tool for qualitative estimates. % Maybe cite the future ISR or LHC TMCI paper \citep{Future_TMCI_paper}.

\subsection{\label{sec:SB}Single-bunch Case}

Let us follow the steps of \citep{bib:NHT} and first consider a single bunch case with a Gaussian beam, represented by a series of nested air-bags. Equation (\ref{eq:NOT_SO_BIG_AND_SCARY}) can be written in a form 
\begin{equation} \label{eq:SB}
\frac{\Delta\omega}{\omega_s} X_{l \alpha} = S_{lm \alpha \beta}X_{m \beta} - iZ_{lm\alpha\beta}X_{m\beta} -igF_{lm\alpha\beta}X_{m\beta},
\end{equation}
where $X$ is the eigenvector of (\ref{eq:NOT_SO_BIG_AND_SCARY}); indexes $l,m$ define azimuthal and $\alpha,\beta$ - radial modes. We will use lower indexes to denote head-tail and upper indexes - coupled-bunch modes.  $S$ denotes the matrix of harmonic oscillations inside the RF potential well:
\begin{equation} \label{eq:S}
S = l\delta_{lm}\delta_{\alpha\beta}.
\end{equation}
It corresponds to the first term in the r.h.s. of Eq.~(\ref{eq:NOT_SO_BIG_AND_SCARY}). The matrix $Z$ defines the single-turn impedance of the problem:
\begin{equation} \label{eq:Z}
Z = i^{l-m}\frac{\kappa}{n_r}\int_{-\infty}^{+\infty}Z_1^\perp(\omega^{\prime}) J_l(\chi^{\prime}_\alpha) J_{m}(\chi^{\prime}_\beta) d\omega^{\prime}.
\end{equation}
Finally, the last term represents the response of a flat damper,  i.e. a damper that acts only on the center of mass of the beam, and whose kick is flat in the time domain. $g$ stands for the normalized damper gain in the units of $\omega_s$. The flat damper is added into Eq.~(\ref{eq:NOT_SO_BIG_AND_SCARY}) as a purely imaginary impedance. Its matrix $F$ can be quickly derived from Eq.~(\ref{eq:Z}) after a substitution $Z(\omega) \approx \delta(\omega)$:

\begin{equation} \label{eq:F}
F = \frac{i^{m-l}}{n_r} J_l(\chi_\alpha) J_{m}(\chi_\beta).
\end{equation}

\subsection{\label{sec:CB_Flat}Multi-bunch Case: Flat Wake}

In the absence of high-Q high-frequency parasitic modes and for sufficiently large bunch spacing the inter-bunch wake can be considered  flat, i.e. its variation over the bunch length can be neglected. In this situation, according to the ``train theorem" \citep{bib:Burov_Transverse_Modes}, all head-tail modes have the same inter-bunch structure. Indeed, a flat wake acts only on the bunch centroid and thus for the inter-bunch motion the bunches are equivalent to macroparticles. This statement is valid for an arbitrary filling pattern of the ring, effectively reducing the coupled-bunch problem to a single-bunch one. The inverse statement is not true though and, generally, the intra-bunch modes depend on the inter-bunch mode number. 

If the inter-bunch wake does not vary significantly over the bunch length it can be simply added into Eq. (\ref{eq:SB}) as an extra term:
\begin{equation} \label{eq:CB}
	\begin{split}
		& \frac{\Delta\omega}{\omega_s} X = S X - iZ X - igF X + C X,\\
        & C_{lm\alpha\beta}^\mu = W^\mu F_{lm\alpha\beta},
	\end{split}
\end{equation}
where $W^\mu$ represents the effect of the inter-bunch wake $W$ on the $\mu$-th coupled-bunch mode. Assuming $M$ equidistant bunches with a bunch spacing of $s_0$, we can write this term as:
\begin{equation} \label{eq:W}
	W^\mu = 2\pi\kappa \sum_{k=1}^{\infty} W(-k s_0)\exp(2\pi i \nu_{k\mu}),
\end{equation}
where the wake $W$ is sampled at the bunch centers with the spacing of $s_0$, $\mu = 0,...M-1$, and $\nu_{k\mu} = k(\mu+\nu)/M$.

Thus for a flat wake the multi-bunch problem reduces to a set of single-bunch problems for each coupled-bunch mode number $\mu$. Furthermore, according to the ``damper theorem" \citep{bib:Burov_Transverse_Modes}, the inter-bunch interaction can be neglected for a sufficiently large damper gain, reducing the multi-bunch problem (\ref{eq:CB}) to the single-bunch case of Eq.~(\ref{eq:SB}).

\section{\label{sec:HOM_Chapter}High frequency HOM}
In the presence of a high-frequency mode the flat wake approximation made above does not hold and one should take into account the variation of the coupled-bunch wake on the bunch length. 
\subsection{\label{sec:HQ_Wake} High-Q resonator wake}
A general high-Q resonator wake can be expressed as 
\begin{equation} \label{eq:resonator_wake}
	\begin{split}
  		& W_1(z) = \frac{cR_s\omega_r}{Q_r}e^{\alpha z/c}\sin \frac{\omega_r z}{c},\\
  		& \alpha = \omega_r/2Q_r,
	\end{split}
\end{equation}
where $R_s$ stands for the shunt impedance, $Q_r$ - quality factor of the mode, and $\omega_r = 2\pi f_r$ - its angular frequency. Denoting $\tilde{W_1}(z)  = \frac{cR_s\omega_r}{Q_r}e^{\alpha z/c}$ and $\phi_k = \omega_r ks_0/c$ one can write the wake of the $k$-th bunch as
\begin{equation} \label{eq:ts_wake}
  W_1(z-ks_0) = \tilde{W_1}(-ks_0)\sin(\frac{\omega_r z}{c} - \phi_k).
\end{equation}
Here we are assuming that $Q_r$ is sufficiently large, such that the decay of the wake over one bunch length can be neglected. This is a valid assumption for HL-LHC where the bunches are much shorter then the bunch spacing. The impedance, corresponding to this wake, consists of two $\delta$-functions at $\pm\omega_r$. Following the logic similar to the flat wake case, described above, we obtain the new coupled-bunch term (the details of derivation can be found in Appendix A)
\begin{equation} \label{eq:CB_term}
  	\begin{split}
    C^\mu & = -\frac{i\pi\kappa}{n_r} \sum_{k = 1}^{\infty} \tilde{W_1}(-ks_0) e^{2\pi i \nu_{k\mu}} i^{m-l}(J^- - J^+), \\
    J^\pm & = J_l(\chi_\alpha \pm \omega_r\tau_\alpha)J_m(\chi_\beta \pm \omega_r\tau_\beta)e^{\pm i\phi_k}.
    \end{split}
\end{equation}
It is easy to see that the final result (\ref{eq:CB_term}) is similar to the flat-wake case (see Eqs.~(\ref{eq:CB},\ref{eq:W})), except the Bessel functions are shifted by $\pm \omega_r$ and an additional phase factor $\phi_k$ appears. 

\subsection{\label{sec:Flat_Wake_Limit}Flat Wake Limit}

If the mode frequency is so low that $\omega_r \ll \omega_\xi$, it can be neglected in the arguments of Bessel functions in (\ref{eq:CB_term}). Then the last term of the product in (\ref{eq:CB_term}) becomes simply $-2i\sin\phi_kJ_l(\chi_\alpha)J_m(\chi_\beta)$. Noting that the wake, created by the $k$-th bunch is $W(-ks_0) = \tilde{W_1}(-ks_0)\sin(-\phi_k)$ we can rewrite the whole coupled-bunch term for the $\mu$-th coupled-bunch mode as:

\begin{equation} \label{eq:CB_term_flat}
  C^\mu = 2\pi\kappa \sum_{k = 1}^{\infty} W(-ks_0)\exp(2\pi i\nu_{k\mu}) F, 
\end{equation}
which is exactly the result obtained above in Eqs.~(\ref{eq:F}-\ref{eq:W}) and matches the coupled-bunch approximation of \citep{bib:NHT}.

\subsection{\label{sec:Weak_HT_Limit}Weak Head-Tail Limit}

A simpler solution can be obtained in the weak head-tail limit when the perturbation is small enough that the mode tune shift is much smaller than $\omega_s$. Then the solution of the linearized Vlasov equation can be approximated by substituting the unperturbed solution. For the air-bag model plugging in ($x_l = \delta_{ll'}, \Omega = \omega_\beta + l\omega_s$) into the right hand side of Eq.~(\ref{eq:NOT_SO_BIG_AND_SCARY}) we obtain a simple formula (\citep{bib:Chao}, Eq.~(6.188)):

\begin{equation} \label{eq:weak_head_tail}
	\frac{\Delta\omega}{\omega_s} = l -i\kappa\sum_{p=-\infty}^{+\infty}Z_1^\perp(\omega') J_l^2(\chi^{\prime}).
\end{equation}

\begin{figure}
\centering
\includegraphics[width = 3in]{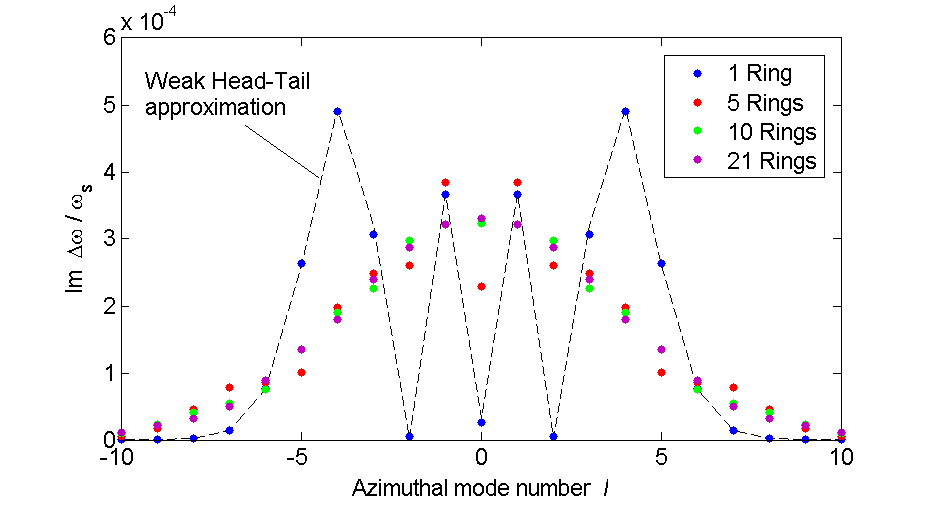}
\caption{
For the airbag distribution, the NHT solution (blue dots) exactly matches the results of the analytical formula for weak head-tail Eq.~(\ref{eq:weak_head_tail}) (dashed line). As we change the distribution to a Gaussian by adding more radial rings the growth rates of the individual azimuthal modes change slightly, but the overall picture remains qualitatively the same: many azimuthal modes get excited and they have similar growth rates. Equispaced beam of 3564~bunches, $10^{10}$~ppb, $\xi = 0$, no damper.
}
\label{fig:conv}
\end{figure}

The simple model in Eq.~(\ref{eq:weak_head_tail}) already gives important insights about the dynamics of collective instabilities, driven by high frequency HOMs. As an example, let us consider the parameters of HL-LHC (summarized in Table~\ref{tab:HL-LHC_params}) with no other sources of impedance $Z(\omega)$ but just one HOM with a frequency $f_{HOM} = 1.97$~GHz and a shunt impedance $R_s = 2.7$~M$\Omega$/m. At $\xi = 0$ we see multiple modes with various azimuthal numbers having substantial growth rates, whereas in the flat wake approximation only the $l = 0$ mode would be excited (Fig.~\ref{fig:conv}).  The weak head-tail approximation precisely matches the exact solution for 1 radial ring, obtained from (\ref{eq:SB}), (\ref{eq:CB}), and (\ref{eq:CB_term}), as seen in Fig.~\ref{fig:conv}. The airbag model features spikes of higher growth rate at certain mode numbers. As we increase the number of radial rings, bringing the longitudinal distribution from an air-bag to a Gaussian (going from 1 to 21 radial rings), the peaks smoothen out. For a Gaussian distribution, mode $l = 0$ has the largest growth rate, but the distribution of growth rates remains wide: the modes with azimuthal numbers as large as 5-10 are excited. The high-$l$ modes are invisible for a flat damper that acts only on $l = 0$ mode at zero chromaticity and a few modes with lower azimuthal numbers at moderately high $\xi$. Thus a flat damper is inefficient for suppressing the coupled-bunch motion, excited by a high frequency HOM.

\begin{figure}
\centering
\includegraphics[width = 3in]{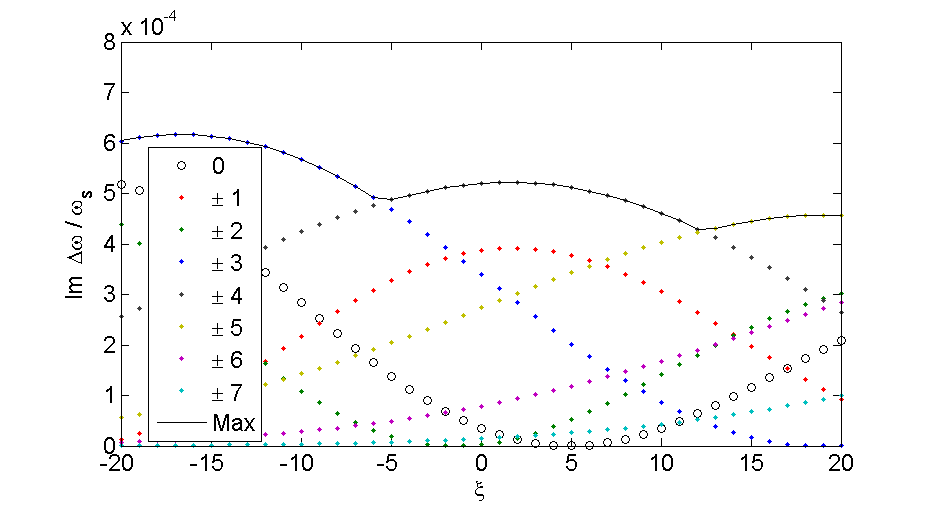}
\caption{
The chromaticity has little effect on the maximum growth rate. Analytic solution for a weak head-tail limit Eq.~(\ref{eq:weak_head_tail}), equispaced beam of 3564~bunches, $10^{10}$~ppb, no damper.
}
\label{fig:modes}
\end{figure}

Since multiple modes with similar growth rates are excited, the growth rate of the most unstable mode does not vary significantly with the chromaticity. Depending on the value of $\xi$ one or another azimuthal mode becomes dominant, but the maximum growth rate remains roughly the same (Fig.~\ref{fig:modes}). Therefore one cannot tune the chromaticity to suppress the coupled-bunch instability.

\begin{table}
  \centering
  \begin{center}
  \caption{Key parameters of HL-LHC, considered for numerical simulation in NHT for the Standard and BCMS (in parenthesis) proton beams}
  \label{tab:HL-LHC_params}
  %\small
    \begin{tabular}{lr}
      \hline\hline
      Parameter & Value \\
      \hline 
Proton energy & 7 TeV \\
Bunch intensity & $2.3\times 10^{11}$~ppb\\
Number of bunches & 2670 (2748)\\
Normalized emittance & $2.1~\mu m$ ($1.7~\mu m$)\\
Revolution frequency & 11.2 kHz\\
Tunes: x, y, z  & 62.31, 60.32, $2.1\times 10^{-3}$\\
Bunch length, rms & 9.0 cm\\
$\beta$* at the end of squeeze & 41 cm\\
Ultimate $\beta$* & 15 cm\\
Feedback damping time & 50 turns\\
\hline 
\hline 
\end{tabular}
%\normalsize
\end{center}
\end{table}

\subsection{\label{sec:PyHHT_Comp} Comparison with tracking simulations}

Another check comes from a comparison with multiparticle tracking simulation. The coupled-bunch simulation was performed in the PyHEADTAIL tracking code \citep{bib:Li, bib:2016grn}. It models the beam as a collection of macroparticles that are symplecticly tracked in 6D. The tracking utilizes a smooth optics approximation and a drift/kick model that treats the accelerator ring as a collection of interaction points, connected by ring segments where the beam is transported via linear transfer maps. The transport maps take into account local Twiss parameters and dispersion at each interaction point but non-linearities such as chromaticity are included indirectly 
as an additional detuning applied to each individual macroparticle.

To include collective effects, such as wake fields, transverse feedback, or space charge, the the beam is longitudinally divided into a set of slices via a 1D particle-in-cell (PIC) algorithm. Collective effects are then applied to macroparticles at the end of each segment on a slice-by-slice basis.

For the comparison we used the HL-LHC beam parameters (Table~\ref{tab:HL-LHC_params}), assuming equidistant bunches for simplicity, and a beam impedance generated by just one HOM, corresponding to the highest mode the 2016 DQW design. The tracking problem included 3564 bunches with $2\times 10^4$ macro-particles per bunch and $1.5\times 10^5$ tracking turns (Table~\ref{tab:PyHHT}). For the purpose of this comparison we assumed an equidistant filling of the whole machine in order to simplify the comparison (In the following Section we scale the growth rate of the coupled-bunch mode with the total intensity to approximate the actual filling scheme). It took $\sim 0.75$ s to simulate one revolution on the CERN HPC cluster with 40 cores. For the purpose of this comparison we did not take into account the Landau damping octupoles, rather focusing on the instability growth rate in the absence of nonlinearities. Simulating the Landau damping by the octupoles would require increasing the number of macroparticles by two orders of magnitude to about $10^6$.

\begin{table}
  \centering
  \begin{center}
  \caption{Parameters of the multibunch PyHEADTAIL tracking simulation used for benchmarking}
  \label{tab:PyHHT}
  %\small
    \begin{tabular}{lr}
      \hline\hline
      Parameter & Value \\
      \hline 
    Proton energy & 7 TeV \\
    Bunch intensity & $2.3\times 10^{11}$~ppb \\
    Number of bunches & 3564\\
    Number of slices per bunch & 20\\
    Macro-particles per bunch & $2 \times 10^4$\\
    Number of tracking turns & $1.5\times 10^5$\\
    Multi-turn wake cut-off & 3 revolutions \\
    Mode frequency & 1.97~GHz\\
    HOM quality factor & $3.1 \times 10^4$ \\
    Mode shunt impedance & 2.7 M$\Omega$/m \\
    $\beta$* & 15 cm\\
    Feedback damping time & 0, 50 turns \\
    Chromaticity range & -20...+20 \\

\hline 
\hline 
\end{tabular}
%\normalsize
\end{center}
\end{table}

\begin{figure}
\centering
\includegraphics[width = 3in]{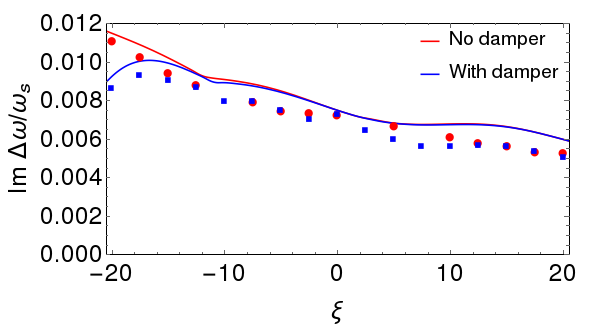}
	\caption{
     NHT solution (solid lines) agrees with the average growth rates found in a multi-bunch particle tracking simulation in PyHEADTAIL (dots). The growth rate created by a single crab cavity HOM varies only slightly with chromaticity and is weakly affected by the presence of a transverse damper. $R_s = 2.7$~M$\Omega$/m, $f = 1.97$~GHz, $Q = 3.1\times 10^4$, $E = 7$~TeV, $\beta^* = 15$~cm, 3564 equidistant bunches of $2.3\times 10^{11}$~p. 
	}
\label{fig:Gr_rate_no_other_sources}
\end{figure}

Both the analytical NHT solution and the tracking simulation show the same qualitative behaviour - the growth rate of the most unstable mode is nearly independent of the transverse damper, except for the lowest chromaticities, and remains similar in a wide range of $\xi$ from -20 to +20 (Fig.~\ref{fig:Gr_rate_no_other_sources}). Quantitatively, the analytical solution is in a good agreement with the tracking, within $\leq 15\%$, both for the case with and without the transverse feedback.

\section{\label{sec:HLLHC} Crab HOMs for HL-LHC}

High-Luminosity LHC employs two sets of crab cavities, each consisting of four cavities per IP and per beam: one set with the crossing in the horizontal and the other - in the vertical plane. Currently, there are two different designs of the crab cavities being considered: Double Quarter Wave (DQW) \citep{bib:Crab_DQW} and RF Dipole (RFD) \citep{bib:Crab_RFD1,bib:Crab_RFD2,bib:Crab_RFD3,bib:Crab_RFD4}. Both designs feature a large number of high order modes, some of which might be dangerous for the coupled-bunch stability. %Table~\ref{tab:CC_HOMs} lists the parameters of the modes with the highest shunt impedance.
For both designs the most dangerous modes have the frequencies of about 2~GHz and shunt impedance of several units of M$\Omega$/m. Initial first-principle estimates indicated that the shunt impedance of the HOMs has to be kept below 1~M$\Omega$/m in order for their effect on the stabilizing octupole current to be small compared to other sources \citep{bib:Crab_Initial_Estimates, bib:Biancacci_2015}. 

We considered the effect of the most critical modes on the transverse beam stability in High-Luminosity LHC, in order to, first, quantify the required increase in the octupole current, and, second, to establish an upper limit on the acceptable mode shunt impedance. To do that we have modified the NHT code \citep{bib:NHT} to include the exact solution of the coupled-bunch problem, discussed in Sec.~IIIA. We used ring impedance and coupled-bunch wakes, obtained with the IW2D code \citep{bib:IW2D}. The code is capable of computing linear driving and detuning wakes. It takes into account various sources of machine impedance, including collimation and injection protection devices, beam screens, pumping ports, etc. 

\begin{figure}
\centering
\includegraphics[width = 3.5in]{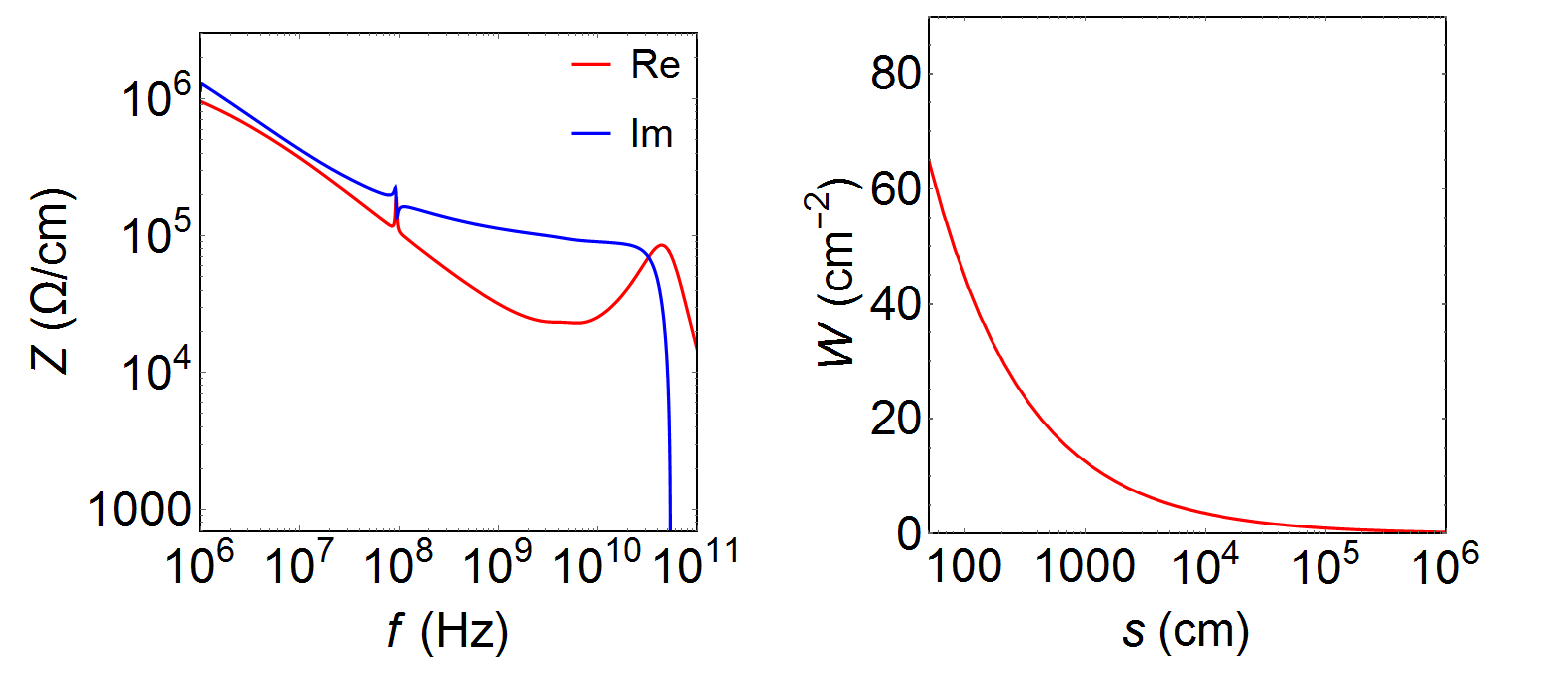}
  \caption{
  Real and imaginary parts of the HL-LHC impedance (left) and the inter-bunch dipolar wake (right) at the top energy before the beams are brought into collision. Crab cavity HOMs not included. Horizontal plane, $\beta^* = 41$~cm, $E = 7$~TeV, Ultimate operational scenario \citep{bib:OP-Note}.
  }
\label{fig:wake}
\end{figure}

For the purpose of this study we first obtained the impedance model of the ring without crab cavities (Fig.~\ref{fig:wake}) and then added the most critical HOMs to the model. Our numerical model incorporated 21 azimuthal modes: from -10 to +10, 9 radial rings, and 19 representative coupled-bunch (CB) modes. The CB modes are chosen such to include the closest ones to the HOM peak, as well as to probe the whole range of CB mode numbers $\mu$ including the one that maximizes the inter-bunch wake $W^\mu$ of the rest of the machine (without the HOMs). The reduction of the number of CB modes allows speeding up the computation without sacrificing the precision: if instead all the CB modes are included, the complex tune shift of the most unstable mode remains the same. 9 radial modes are enough to keep the phase advance between the neighbors, shown in Eq.~(\ref{eq:crit}), small for the frequencies in question. To be confident in the solution, its numerical convergence has been checked on several larger bases of up to 21 radial rings and the relative error was found to be within 5\%.   

\begin{figure}
\centering
\includegraphics[width = 2.5in]{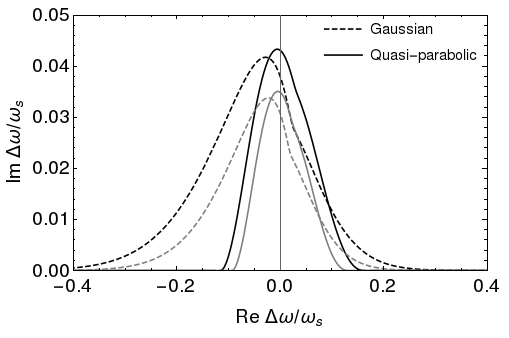}
	\caption{
    The stability diagram for a quasi-parabolic distribution (with the tails cut at $3.2\sigma_{rms}$) is narrower than the one for the Gaussian profile (with infinitely long tails), leading to a higher octupole threshold for the modes that have a large negative real tune shift. Stability diagrams computed for $E = 7$~TeV, $\beta^* = 41$ cm and maximum negative octupole current of $I_{oct} = -570$~A, Standard $\epsilon_n = 2.1~\mu$m (black) and BCMS $\epsilon_n = 1.7~\mu$m (gray) rms beam emittances.
	}
\label{fig:SD}
\end{figure}

Landau damping in HL-LHC is provided by a system of dedicated octupole magnets. The octupoles operate at the negative polarity, which is preferable for the beam lifetime (see for example \citep{bib:Pieloni}, a detailed discussion on the sign of the octupole current can be found in \citep{bib:OP-Note}). At the maximum current of 570~A they produce an rms tune spread of $\delta \nu_{rms} \simeq 10^{-4}$ for the standard beam with $\epsilon_n = 2.1~\mu$m. The corresponding stability diagram depends on the transverse distribution, since the tune shift is produced by the high-amplitude tails. As a conservative estimate, we assume that the tails of the distribution are cut at $\sim 3\sigma_{rms}$ by the collimation system (e.g. in case a hollow electron lens is installed to clean the transverse tails). In this case the modes with a large real frequency shift (typical for a resistive wall type of impedance that is dominant for LHC) would require a larger octupole current to stabilize compared to a normal distribution. The stability diagram produced by this `quasi-parabolic' distribution \cite{bib:SD} is depicted in Fig.~\ref{fig:SD}.

\begin{figure}
\centering
\includegraphics[width = 3in]{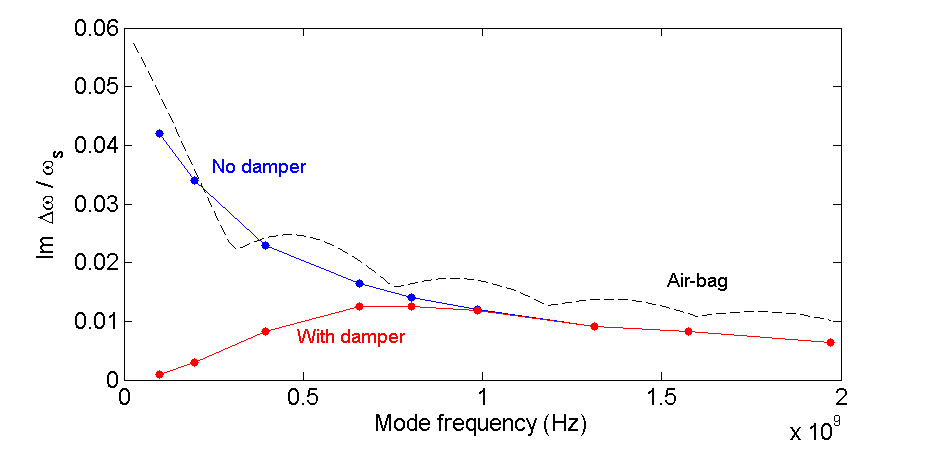}
  \caption{
  At large frequencies, above 1 GHz, the instability growth rate is not affected by the transverse damper. NHT simulation for a Gaussian beam with a 50 turn (blue line) and without (red) the damper and an analytical solution for an air-bag beam in the weak head-tail limit (dashed black line).
  }
\label{fig:freq_scan}
\end{figure}

The transverse damper is inefficient at damping the high frequency HOMs, in particular the ones above 1~GHz (Fig.~\ref{fig:freq_scan}). The HOMs above this limit require special attention. Both crab cavity designs feature a large number of HOMs above 1~GHz that might be potentially dangerous. Table~\ref{tab:CC_HOMs} lists the most dangerous modes of the most up-to-date designs, Tables~\ref{tab:CC_DQW_2016},\ref{tab:CC_RFD_2016} in Appendix B provide the lists of high frequency HOMs from 2016 that ignited this study.

The crab cavity impact on the machine impedance increases linearly with the ratio of $\beta$-functions at its location to the average $\beta$-function in the ring $\beta_{cc}/\beta_{avg}$. At the maximum $\beta_{cc}$, corresponding to the $\beta$-function at the interaction point $\beta^* = 15$~cm and $\beta_{cc}/\beta_{avg} = 52$, the instability growth rate produced by the crab cavities alone can be rather large. As seen in Fig.~\ref{fig:Gr_rate_no_other_sources} a single HOM could produce a growth rate Im~$\Delta\omega/\omega_s \sim 5\times 10^{-3}$ in the absence of any other sources of impedance (Fig.~\ref{fig:Gr_rate_no_other_sources}). Such a large mode growth rate would require around 70~A of octupole current to suppress it. In HL-LHC the beams are brought into collision at $\beta^* = 64$~cm in the Nominal or at $\beta^* = 41$~cm in the more challenging, Ultimate operational scenario \citep{bib:OP-Note}.

\begin{center}
\begin{figure}
\includegraphics[width = 7in]{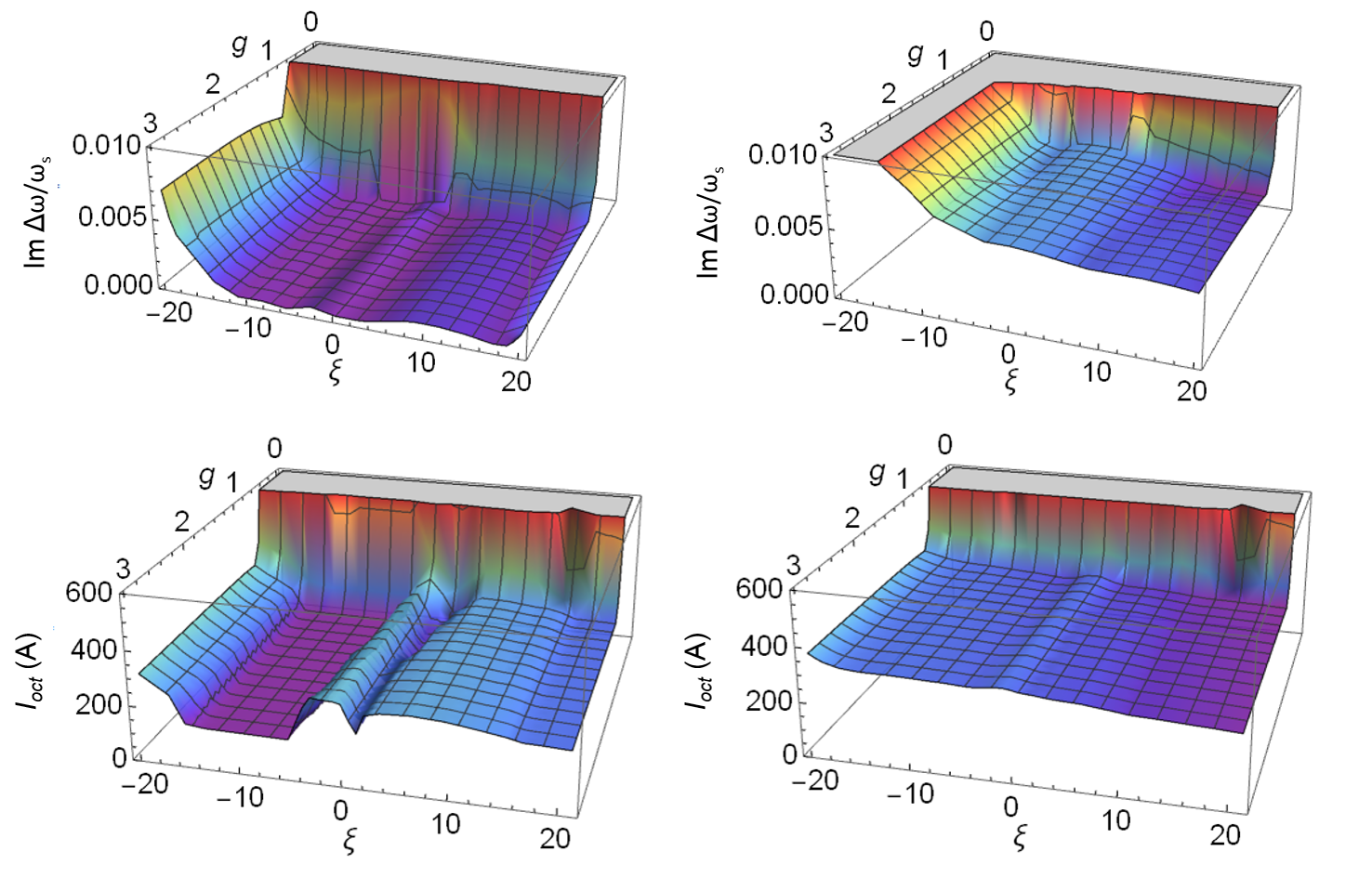}
      \caption{
      Compared to the case without the crab HOMs (left), the coupled-bunch dynamics with one dominant HOM (right) depends significantly weaker on the damper gain (g) or the chromaticity ($\xi$). The growth rate of the most unstable coupled-bunch mode is plotted in the top and the corresponding stabilizing octupole current - in the bottom. Horizontal plane, $R_s = 2.7$~M$\Omega$/m, $f_r = 1.97$~GHz, $\beta^* = 15$~cm, $E = 7$~TeV, $2.3\times 10^{11}$~ppb, $\epsilon_n = 2.1~\mu$m.
      }
  \label{fig:3d_gr_rates}
\end{figure}
\end{center}

We focused on the most critical case for single-beam stability, just before the beams are brought into collision, at the beginning of the luminosity leveling process, when $\beta^* = 41$~cm (for the ultimate luminosity of $7.5\times 10^{34}$~cm$^{-2}$s$^{-1}$) and has not yet reached its minimum value of 15 cm. For the Ultimate scenario $\beta_{cc}/\beta_{avg} \approx 20$. At this ratio, a hypothetical HOM with a sufficiently large $R_s \sim 10$~M$\Omega$/m and sufficiently high $f \sim 2$~GHz can significantly increase the instability growth rate in a wide range of chromaticities and damper gains (Fig. \ref{fig:3d_gr_rates}, top). The resulting growth rate with the HOM is largely independent of both $\xi$ and $g$, indicating their poor effectiveness (as discussed in Sec.~IIIC). Since both the damper and the chromaticity have little effect on the resulting growth rate, beam stability has to be provided through Landau damping.

\begin{table}
  \centering
  \begin{center}
  \caption{Parameters of the most critical crab cavity HOMs for HL-LHC from numerical simulations}
  \label{tab:CC_HOMs}
  %\small
    \begin{tabular}{lrr}
      \hline\hline
      Crab cavity type & RFD & DQW\\
      \hline 
Mode frequency & 1.81 GHz & 1.92 GHz\\
Quality factor & 20000 & 59000\\
Shunt impedance & 1.0 M$\Omega$/m & 2.2 M$\Omega$/m \\

\hline 
\hline 
\end{tabular}
%\normalsize
\end{center}
\end{table}

\onecolumngrid
\twocolumngrid

In the absence of the high frequency HOMs around 250~A is required to ensure the transverse beam stability in HL-LHC (Fig.~\ref{fig:3d_gr_rates}, bottom-left). That leaves a factor of two safety margin to account for possible linear and nonlinear lattice errors, coupling, long-range beam-beam interaction, and uncertainties in beam distribution and impedance model. Such a safety margin has proven to be sound in LHC operation. It is important that the margin is not reduced significantly by the addition of the crab cavity HOMs, thus their impedance has to be kept under control. If the shunt impedance is too high the octupole current might increase dramatically in the whole operational range of chromaticities and damper gains (Fig.~\ref{fig:3d_gr_rates}, bottom-right). In that case it would be nearly impossible to improve beam stability by tuning in either of the settings.

\begin{figure}[]
\centering
\includegraphics[width = 3.0in]{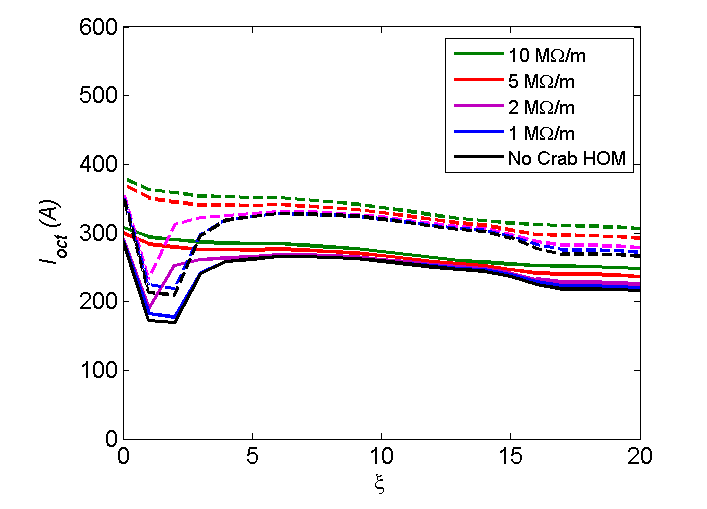}
	\caption{
	In order not to affect the HL-LHC operational scenarios the HOM shunt impedance has to be kept below $\sim$1~M$\Omega$/m. NHT simulation for the top energy of $E = 7$~TeV; Ultimate operational scenario, $\beta^* = 41$~cm, Standard (solid lines) and BCMS (dashed lines) beam parameters.  
	}
\label{fig:Oct_current_new_OP}
\end{figure}

When setting a limit on the maximum acceptable HOM shunt impedance one should take into account that due to manufacturing uncertainties the quality factors $Q$ of some crab cavity HOMs are likely to vary from the design values and from cavity to cavity. That, in turn, might lead to a variation of the shunt impedance, since the two are related as $R_s = Q\times R/Q$. According to the recent measurements at CERN \citep{bib:J_Mitchell_Madrid}, the actual $Q$ may be higher than the simulated one by up to a factor of three for some modes. On the other hand, thanks to the manufacturing uncertainties the HOMs of different cavities are unlikely to overlap. A 3\% rms deviation of mode frequencies from the table values has been measured in the test cavities \citep{bib:J_Mitchell_Madrid}, which corresponds to 3~MHz uncertainty of a mode frequency of 1~GHz - much larger than the typical width of the HOMs. For the most challenging HL-LHC operational scenario, where the beams are brought into collision at $\beta^* = 41$~cm, in order for the crab cavity not to affect the beam stability significantly their $R_s$ should be kept below 1~M$\Omega$/m. In that case the additional increase of the octupole current shall not exceed 10~A in the operational range of chromaticities $\xi = 5-15$ (Fig.~\ref{fig:Oct_current_new_OP}).

\section{\label{sec:Conclusion}Conclusion and outlook}

In order to assess the impact of crab cavities on the transverse stability in HL-LHC we have extended the nested airbag approach of Ref.~\citep{bib:NHT} to describe fast oscillating coupled-bunch wakes of the crab cavity HOMs and implemented it in the Nested Head-Tail Vlasov solver code. NHT includes radial and coupled-bunch modes, feedbacks, and Landau damping (in the weak head-tail regime) and allows performing fast and computationally efficient multidimensional parameter scans.

The analysis shows that a single HOM from one HL-LHC crab cavity may require a significant amount of octupole current to be Landau damped even in the absence of other sources of impedance. In order for the HOMs not to limit the choice of chromaticity setting, the associated increase of the octupole threshold has to remain below 10~A in a wide range of positive chromaticites: both as high as $\sim~15$ (current LHC setting) and as low as $\sim~5$ (in case of future needs) for the most critical operational scenario \citep{bib:OP-Note}. This requirement is satisfied if no HL-LHC crab cavity HOM exceeds a transverse shunt impedance limit of 1~M$\Omega$/m, otherwise the mode's impact on beam stability for particular operational parameters has to be examined separately.

It has to be noted that the present analysis considers the Ultimate operational scenario where the two beams are brought into collision at $\beta^* = 41$~cm (for levelling at a peak luminosity of $7.5\times 10^{34}$~cm$^{-2}$s$^{-1}$). The Nominal operational scenario features a 1.5 times larger $\beta^*$ at the start of the collision process ($\beta^* = 64$~cm, peak luminosity of $5\times 10^{34}$~cm$^{-2}$s$^{-1}$), proportionally lowering the impact of crab cavities on the transverse dynamics. Even in the very unlikely scenario that the machine cannot be operated with the crab cavities, its performance could be partially restored with alternative techniques such as flat beam optics and beam-beam wire compensation \citep{bib:2015djt, bib:2015ais}. The flat optics (not in the HL-LHC baseline at the moment) is estimated to provide 214~fb$^{-1}$ of integrated luminosity without and 225~fb$^{-1}$ of with the compensation, compared to the 240~fb$^{-1}$ for the Nominal operational scheme with crab cavities and the same effective event pile-up density \citep{bib:HL-LHC_TDR}.

Thanks to a dedicated effort to suppress the HOMs over recent years, most HOMs of both HL-LHC crab cavity designs (RFD and DQW) are expected to be significantly lower than 1~M$\Omega$/m with individual ones close to the limit. Recent measurements of a DQW prototype in CERN SPS confirm HOM transverse shunt impedances, predicted in simulation \citep{bib:J_Mitchell_CERN}. The SPS test did not allow directly testing the destabilizing effect of transverse HOMs due to low $\beta$-functions at the crab cavities, instead what was measured was the power load on the HOM couplers. The measurements revealed that no prominent modes were missed in numerical simulation and gave a valuable input of HOM frequency and quality factor tolerances which have been taken into account in this study. The test also identified the modes that need to be closely monitored during cavity production and testing. A few HOMs in the SPS cavities exceed 1~M$\Omega$/m with the largest around 5~M$\Omega$/m, which can be attributed to a different design of the couplers dictated by spatial constraints in SPS and their 5-mm misalignment from the nominal positions. The HL-LHC HOM coupler design already incorporates several improvements not implemented in SPS and further optimization is currently ongoing to meet mechanical and cooling constraints. 

High frequency HOMs can be potentially dangerous for future circular colliders that take advantage of crab crossing scheme. The transverse higher order modes of crab cavities may drive coupled-bunch modes with a fast rise time as the $\beta$-function at the interaction point is squeezed. Due to their high frequency the HOMs typically drive multiple intra-bunch modes with high azimuthal numbers, making both the chromaticity and a flat resistive feedback inefficient against the resulting instability. The effect is stronger the smaller the $\beta^*$ at which the beams are brought into collision (meaning larger the $\beta$-functions at the cavities) and may require additional means of beam stabilization or even limit machine performance. For that reason crab cavity HOM shunt impedance has to be kept under control to ensure coherent beam stability. Related impedance reduction studies are envisioned for example, for HE-LHC \citep{bib:HE-LHC_CDR}.

\section*{Acknowledgments}

This research is supported by the High-Luminosity LHC project. The authors would like to thank Kevin Li for his help with multibunch particle tracking simulations in PyHEADTAIL, Rama Calaga and James Mitchell for providing simulation and measurement data on the HL-LHC crab cavity HOMs, and Gianluigi Arduini for discussion and comments.

\newpage
%%%%%%%%%%%%%%%%%%%%%%%%%%%%%%%%%%%%%%%%%%%%%%%%%%%%%%%%%%%%%%%%%%%%%%%%
\appendix
%%%%%%%%%%%%%%%%%%%%%%%%%%%%%%%%%%%%%%%%%%%%%%%%%%%%%%%%%%%%%%%%%%%%%%%%

\section{\label{sec:Exact_Sol} Exact solution for a sinusoidal wake}

In this Appendix we derive the exact solution for the high-frequency coupled-bunch wake. Let us start from the linearized Vlasov equation (\ref{eq:Linearized_Vlasov}) and consider the transverse deflecting force, acting on a bunch. The transverse wake force $F_y$, generated by the dipole moment of the beam $D$, can be written as (see \citep{bib:Chao}, Eq.~6.239):
\begin{equation} \label{eq:Fy}
	\begin{split}
	& F_y(z,s) = -\frac{e^2 D}{c T_0} \sum_{k=-\infty}^{+\infty}\int_{-\infty}^{+\infty} dz^{\prime}\sum_{n=0}^{M-1} e^{\frac{2\pi in\mu}{M}} \times \\ 
    & \times W_1(z-z^{\prime}-kC-\frac{nC}{M}) e^{-i\frac{\Omega}{c}(s-kC-\frac{nC}{M})} \rho(z^{\prime}) ,
    \end{split}
\end{equation}
where $C$ stands for the ring circumference. The first summation in $k$ goes over all turns, and the second one, in $n$, - over all bunches in the ring. $n=0$ corresponds to a single bunch wake, and $n\geq 1$ - interaction with the other bunches. Using the additivity of the wakes we can consider the coupled-bunch sum separately. Replacing the double summation with one we obtain:
\begin{equation} \label{eq:Fy1}
	\begin{split}
  		F^{CB}_y(z,s) = & -\frac{e^2 D}{c T_0} e^{-\frac{i\Omega s}{c}} \sum_{k=1}^{+\infty} e^{2\pi i\nu_{k\mu}} \times \\
        & \times \int_{-\infty}^{+\infty} W_1(z-z^{\prime}-ks_0) \rho(z^{\prime})dz^{\prime}.
	\end{split}
\end{equation}
In the frequency domain the time-shifted wake $W_1(z-ks_0)$ corresponds to an impedance, multiplied by an additional phase factor:
\begin{equation} \label{eq:Zn}
  \frac{i}{c}\int_{-\infty}^{+\infty}W_1(z-ks_0)e^{-i\omega z/c}dz = Z_1^\perp(\omega)e^{-i\omega ks_0/c}.
\end{equation}

We now obtain the solution for an equispaced air-bag beam with a radius $c\tau$ and bunch spacing $s_0$ by substitution of Eqs.~(\ref{eq:Fy1},\ref{eq:Zn}) into (\ref{eq:NOT_SO_BIG_AND_SCARY}). The right hand side then becomes
\begin{equation} \label{eq:RHS}
	\begin{split}
		r.h.s. = &lx_l -i\kappa \sum_{l'=-\infty}^{+\infty} x_{l'}i^{l-l'} \sum_{k=1}^{+\infty} e^{2\pi i \nu_{k\mu}} \times \\
  		& \times \sum_{p=-\infty}^{+\infty}Z_1^\perp(\omega') e^{-i\omega' ks_0/c} J_l(\chi') J_{l'}(\chi'),
	\end{split}
\end{equation}
or, in terms of Eqs.~(\ref{eq:SB},\ref{eq:CB}) 
\begin{equation} \label{eq:CBnew}
	\frac{\Delta\omega}{\omega_s}X = S_{lm}X + C_{lm}^{\mu}X.
\end{equation}
Note, in this derivation we omitted single-bunch and damper terms.

For a high-Q resonator its wake and impedance can be written as:
\begin{equation} \label{eq:highQ}
	\begin{split}
		W_1(z) & = W_r \sin(\omega_r z/c), \\
  		Z_1^\perp(\omega) & = \pi W_r(\delta(\omega-\omega_r)+\delta(\omega+\omega_r)),
	\end{split}
\end{equation}
where $W_r = cR_s/Q$ is the magnitude of the wake, $\omega_r$ - its angular frequency, $R_s$ - shunt impedance, and $Q$ - quality factor. Substituting (\ref{eq:highQ}) into (\ref{eq:RHS}, \ref{eq:CBnew}) we obtain
\begin{equation} \label{eq:RHS_highQ}
	\begin{split}
		C_{lm}^{\mu} = & -i\pi\kappa W_r i^{l-m} \sum_{k=1}^{+\infty} e^{2\pi i \nu_{k\mu}} (J^{-} - J^{+}), \\
  		J^{\pm} = & J_l(\chi \pm \omega_r\tau)J_{m}(\chi \pm \omega_r\tau)e^{\pm i\phi_k},
    \end{split}
\end{equation}
where the phase factor $\phi_k = \omega_r k s_0/c$.

\section{\label{sec:HOM_Tables} HOM tables}

In this section we provide the most critical HOMs for the two HL-LHC designs: DWQ (Table~\ref{tab:CC_DQW_2016}) and RFD (Table~\ref{tab:CC_RFD_2016}). For the sake of conciseness only the modes with transverse shunt impedance of or larger than 0.1~M$\Omega$/m are presented.

\begin{table}[h]
  \centering
  \begin{center}
  \caption{DQW HOMs with the largest transverse shunt impedance. Data of 2016}
  \label{tab:CC_DQW_2016}
  %\small
    \begin{tabular}{lccr}
      \hline\hline
      $f$, GHz & Plane & $Q$ & $R_s$, M$\Omega$/m \\
      \hline 
    1.977 & Hor. & $3.1 \times 10^4$ & 2.7 \\
    1.50 & Hor. & $3.0 \times 10^4$ & 1.2 \\
    0.681 & Hor. & $6.7 \times 10^3$ & 1.0 \\
    1.960 & Vert. & $5.0 \times 10^4$ & 0.2 \\
    1.726 & Vert. & $2.2 \times 10^4$ & 0.2 \\
    1.656 & Hor. & $7.1 \times 10^3$ & 0.2 \\
    0.928 & Hor. & $610$ & 0.2 \\
    1.747 & Vert. & $1.4 \times 10^3$ & 0.1 \\
    1.744 & Hor. & $1.0 \times 10^4$ & 0.1 \\
\hline 
\hline 
\end{tabular}
%\normalsize
\end{center}
\end{table}

\begin{table}
  \centering
  \begin{center}
  \caption{RFD HOMs with the largest transverse shunt impedance. Data of 2016}
  \label{tab:CC_RFD_2016}
  %\small
    \begin{tabular}{lccr}
      \hline\hline
      $f$, GHz & Plane & $Q$ & $R_s$, M$\Omega$/m \\
      \hline 
    1.960 & Hor. & $2.4 \times 10^5$ & 4.2 \\
    2.0 & Hor. & $1.1 \times 10^5$ & 4.2 \\
    1.99 & Hor. & $6.6 \times 10^4$ & 3.3 \\
    2.042 & Hor. & $2.0 \times 10^5$ & 0.9 \\
    1.960 & Hor. & $5.6 \times 10^4$ & 0.8 \\
    1.770 & Hor. & $1.1 \times 10^4$ & 0.8 \\
    0.634 & Hor. & $1.4 \times 10^4$ & 0.7 \\
    1.730 & Vert. & $4.0 \times 10^4$ & 0.6 \\
    1.270 & Vert. & $2.6 \times 10^3$ & 0.6 \\
    1.475 & Hor. & $7.6 \times 10^3$ & 0.4 \\
    1.878 & Hor. & $1.3 \times 10^5$ & 0.3 \\
    1.481 & Vert. & $1.6 \times 10^3$ & 0.3 \\
    1.990 & Hor. & $1.2 \times 10^4$ & 0.2 \\
    1.972 & Vert. & $1.8 \times 10^4$ & 0.1 \\
\hline 
\hline 
\end{tabular}
%\normalsize
\end{center}
\end{table}

\end{document}